\documentclass[pre,floatfix,a4paper,twocolumn,nofootinbib,amssymb,
               amsmath,nopacs,showkeys,superscriptaddress]{revtex4}
\usepackage{graphicx}

\newcommand {\be} {\begin{equation}}
\newcommand {\ee} {\end{equation}}
\newcommand {\Be}{\begin{eqnarray*}}
\newcommand {\Ee} {\end{eqnarray*}}
\newcommand {\bey} {\begin{eqnarray}}
\newcommand {\eey} {\end{eqnarray}}
\newcommand{\bit}{\begin{itemize}}      
\newcommand{\eit}{\end{itemize}}
\newcommand{\bfl}{\begin{flusleft}}
\newcommand{\efl}{\end{flusleft}}
\newcommand{\bfr}{\begin{flushright}}
\newcommand{\bc}{\begin{center}}
\newcommand{\ec}{\end{center}}
\newcommand{\ben}{\begin{enumerate}}    
\newcommand{\een}{\end{enumerate}}

\newcommand{\comment}[1]{}

\newcommand{\av}[1]{\left\langle #1\right\rangle}

\begin{document}

\title{Desynchronized stable states in diluted neural networks} 
\author{R\"udiger Zillmer}
\email{zillmer@fi.infn.it}
\affiliation{INFN Sez.~Firenze,
via Sansone, 1 - I-50019 Sesto Fiorentino, Italy}
\author{Roberto Livi}
\email{livi@fi.infn.it}
\affiliation{Dipartimento di Fisica - CSDC, Universit\'a di Firenze,
via Sansone, 1 - I-50019 Sesto Fiorentino, Italy}
\affiliation{INFN Sez.~Firenze,
via Sansone, 1 - I-50019 Sesto Fiorentino, Italy}
\author{Antonio Politi}
\email{antonio.politi@isc.cnr.it}
\affiliation{Istituto dei Sistemi Complessi, 
CNR, via Madonna del Piano 10, I-50019 Sesto Fiorentino, Italy}
\author{Alessandro Torcini}
\email{alessandro.torcini@isc.cnr.it}
\affiliation{Istituto dei Sistemi Complessi, 
CNR, via Madonna del Piano 10, I-50019 Sesto Fiorentino, Italy}
\affiliation{INFN Sez.~Firenze,
via Sansone, 1 - I-50019 Sesto Fiorentino, Italy}

%%%%%%%%%%%%%%%%%%%%%%%%%%%%%%%%%%%%%%%%%%%%%%%%%%%%%%%%%%%%%%%%%%%%%%%%%%%
\begin{abstract}
The dynamical properties of a diluted fully-inhibitory network of 
pulse-coupled neurons are investigated. Depending on the coupling strength,
two different phases can be observed. At low coupling the evolution rapidly
converges towards periodic attractors where all neurons fire with the same rate.
At larger couplings, a new phase emerges, where all neurons are mutually
unlocked. The irregular behaviour turns out to be ``confined" to an
exponentially long, stationary and linearly stable transient. In this latter
phase we also find an exponentially tailed distribution of the inter-spike intervals
(ISIs). Finally, we show that in the unlocked phase a subset
of the neurons can be eventually synchronized under the action
of an external signal, the remaining part of the neurons acting as a background
noise. The dynamics of these ''background'' neurons is quite peculiar, in that
it reveals a broad ISI distribution with a coefficient of variation that
is close to 1.
\end{abstract} 

%%%%%%%%%%%%%%%%%%%%%%%%%%%%%%%%%%%%%%%%%%%%%%%%%%%%%%%%%%%%%%%%%%%%%%%%%%
% 87.19.La - Neuroscience
% 84.35.+i Neural networks
% 05.45.-a Nonlinear dynamics and nonlinear dynamical systems
% 05.45.Xt Synchronization; coupled oscillators
% 87.10.+e General theory and mathematical aspects (biophysics)
% 07.05.Mh Neural networks, fuzzy logic, artificial intelligence
% 89.75.-k Complex systems
\maketitle
%%%%%%%%%%%%%%%%%%%%%%%%%%%%%%%%%%%%%%%%%%%%%%%%%%%%%%%%%%%%%%%%%%%%%%%%%%%%%
\section{ Introduction}
A challenging task in the context of complex neural networks is the
comprehension of how their dynamical features are influenced by their structure.
A good strategy to tackle the problem consists in studying simple models, since
one can more easily identify the minimal and relevant ingredients that are
responsible for generic properties. In this work we investigate a diluted
neural network of pulse--coupled neurons. \footnote{Here dilution refers to a random
pruning of directed links in a globally coupled network.}
Since inhibition plays an important
role in the dynamics of neurons in vivo\cite{rudolph,brunel-wang}, we have chosen
to examine a network of inhibitory coupled leaky integrate--and--fire neurons. 
Recently, it has been shown that dilution can induce long chaotic transients in
excitatory networks with delay \cite{Zumdieck-Timme:2004}. We aim at
understanding the new features emerging in the dynamics of a network where two
competitive effects like dilution and full inhibition coexist. We show that the
maximum Lyapunov exponent is negative \footnote{Strictly speaking, it is the
second exponent, as there exists an exponent equal to zero that corresponds to
the motion along the trajectory. For the sake of clarity and since we study the
dynamics emerging from the introduction of a Poincar\'e section, we always
disregard the zero exponent.}, so that the evolution eventually
converges onto a periodic attractor. This notwithstanding, the dynamics is
far from trivial, because above a given threshold, the transient grows
exponentially with the system size and is characterized by seemingly statonary
properties. These features altogether, plus the stochastic-like nature of the
transient, suggest a strong analogy with {\sl stable chaos} (SC),
a phenomenon discovered in the context of coupled map lattices \cite{stable}. 
The possibility of observing SC in the neuroscience context is important because
this regime is both stable to external perturbations and characterized by a
richness of behaviour depending on the initial conditons: both properties are
welcome for a system that is expected to reliably perform universal
computations. The present case study can be considered as the first
example of SC in an autonomous continuous-time system. In fact, it was formerly
believed that SC can arise only in periodically forced systems, as one must
simultaneously have a zero Lyapunov exponent (in order to ensure a
meaningful dynamics different from a dull fixed point) and a strictly negative
second Lyapunov exponent. The continuity of Lyapunov spectra in autonomous
spatially extended systems prevents the above condition to be fulfilled and, in
fact, the only previous SC example in a continuous time system is a chain of
periodically forced Duffing oscillators \cite{bonac}. However, these arguments do
not apply to globally coupled systems and this explains why SC can and has been
observed in neural networks.

%%%%%%%%%%%%%%%%%%%%%%%%%%%%%%%%%%%%%%%%%%%%%%%%%%%%%%%%%%%%%%%%%%%%%%%%%%%%%

\section{Dynamics of a weakly disordered network}
We consider a system made of $N$ leaky integrate-and-fire neurons, which
interact by sending each other pulses through a directed graph of
connections. The non-zero elements of the connectivity matrix are
$G_{ij}=G_0/\ell_i\,$, where $G_0$ is the coupling constant and $\ell_i$ is the 
number of incoming links to neuron $i\,$. Such a normalization condition has
been used in the literature when dealing with randomly connected
networks \cite{Zumdieck-Timme:2004,Gerstner-Kistler:2002}.
The membrane potential $v_i(t)$ of the $i$-th neuron obeys the dynamical rule
\be\label{eq:model1}
  \dot{v}_{i}=c-v_{i}-(v_i+w)\,
  \sum_{j=1}^{N}\sum_{m=1}^{M_j}G_{ij}
  \,\delta(t-t_j^{(m)})\,\,\, ;\,\,\,
\ee  
where we use dimensionless quantities (details about the derivation of the model
can be found in \cite{zillmer}). The suprathreshold input current, $c$, 
and the reversal potential, $w$, are assumed to take the values 2 and 4/7,
respectively.  This choice was derived from values used in the current
literature (see e.g., \cite{Brunel-Hakim:1999,Jin:2002}).
The potential $v_i(t)$ follows the above dynamics until it reaches the threshold
value $1$, whereupon the neuron emits a spike and the potential is reset to the
value $0$. The integer $M_j$ in the coupling term counts the spikes
emitted by neuron $j$ at the time $t_j^{(m)}$. Since we assume
$G_0>0$ and $w > 0$, the coupling turns out to be fully inhibitory.

One further important ingredient of the model is {\it disorder} that is introduced
by randomly pruning a fraction $r_{\rm p}$ of directed links. As a matter of
fact, the main control parameters are $r_{\rm p}$ and the coupling
constant $G_0$. 

By introducing a Poincar\'e section (defined as the manifold where a generic
neuron potential reaches the threshold), it has been shown that the network
dynamics can be reduced to a discrete map of the potentials $v_i(n)$ between
successive spikes \cite{zillmer}. Accordingly, one can replace the continuous
time axis with an integer $n$ denoting the $n$th spike. It is convenient to
simulate the model by directly iterating the resulting map.

A preliminary analysis has been performed in Ref.~\cite{Jin:2002}, where it was
shown that the evolution converges to a periodic state and rigorous upper bounds
for the transient time have been thereupon derived. Accordingly, the dynamics
seems to be quite straightforward and in fact the maximum Lyapunov exponent
remains negative even in the thermodynamic limit \cite{zillmer}. More precisely, a
homogeneous fully coupled network rapidly converges towards a periodic phase
with all neurons firing with the same rate. On the other hand, the transient 
length may
become exponentially long with the system size, so that in large networks the
name {\it transient} becomes inadequate to identify what turns out to be a
long-lasting stationary regime. Interestingly enough, upon increasing the
amount of disorder (i.e., by increasing the fraction $r_{\rm p}$ of cut links)
one passes from the former to the latter regime. The results reported in this
section for $r_{\rm p}$ as small as 0.05, show that for sufficiently large
coupling strengths a stochastic-like behaviour is self-generated.

At low coupling strengths (up to $G_0 \alt 1$), the
evolution is substantially equivalent to that of a fully coupled system. After
a short transient time, on the order of the system size $N$, all neurons fire
with the same pace ($t_{\rm isi} = const$) but are characterized by different
absolute phases. The main difference between this synchronized regime, that we
call {\sl Locked Phase} (LP), and that one occurring in the fully coupled case
is the difference among the relative phases. In fact, the equivalence
among all neurons of a homegeneous network forces the evolution to converge
towards a highly symmetric state characterized not only by a constant
interspike interval (ISI) $t_{\rm isi}$, but also by a constant time separation
between
consecutive spikes. On the other hand, already the introduction of a small
amount of disorder breaks the symmetry, and different spiking sequences are no
longer equivalent to one another, with the consequence that slightly different
time series are generated for different initial conditions.
Although this phase is dynamically trivial, it may have some relevance from the
point of view of information storage, since there is an
exponentially large (and even more than that) number of different attractors
that may be selected by means of suitable initial inputs.

\begin{figure}[h]
\begin{center}
\includegraphics[width=0.37\textwidth,clip=true]
{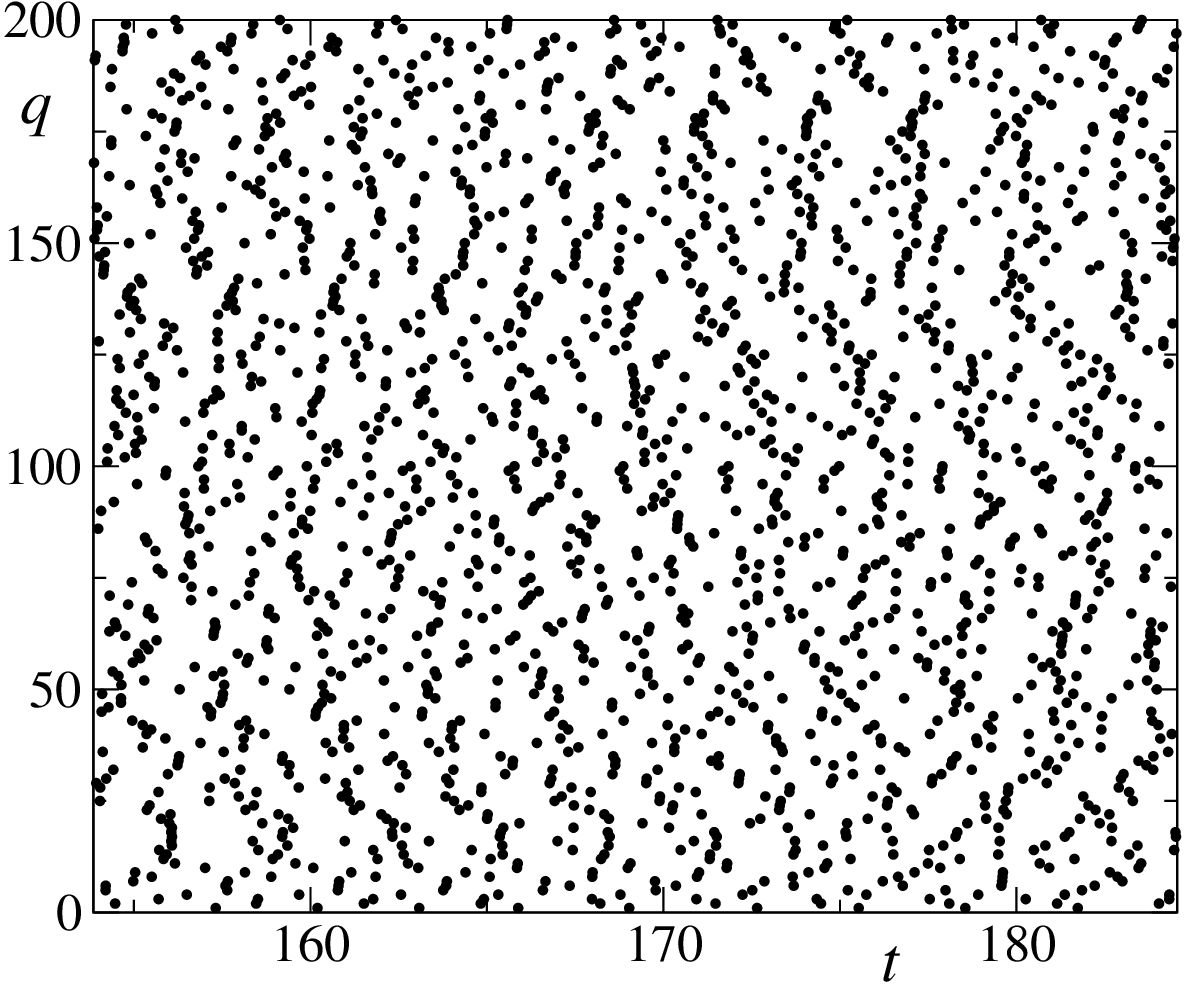}
\caption{Firing pattern during the transient in the UP ($G_0=2,N=200,r_{\rm p}
=0.05\,$).
$q$ denotes the index of the firing neuron.}
\label{fig:perpat1}
\end{center}
\end{figure}

\begin{figure}[h]
\includegraphics[width=0.35\textwidth,clip=true]
{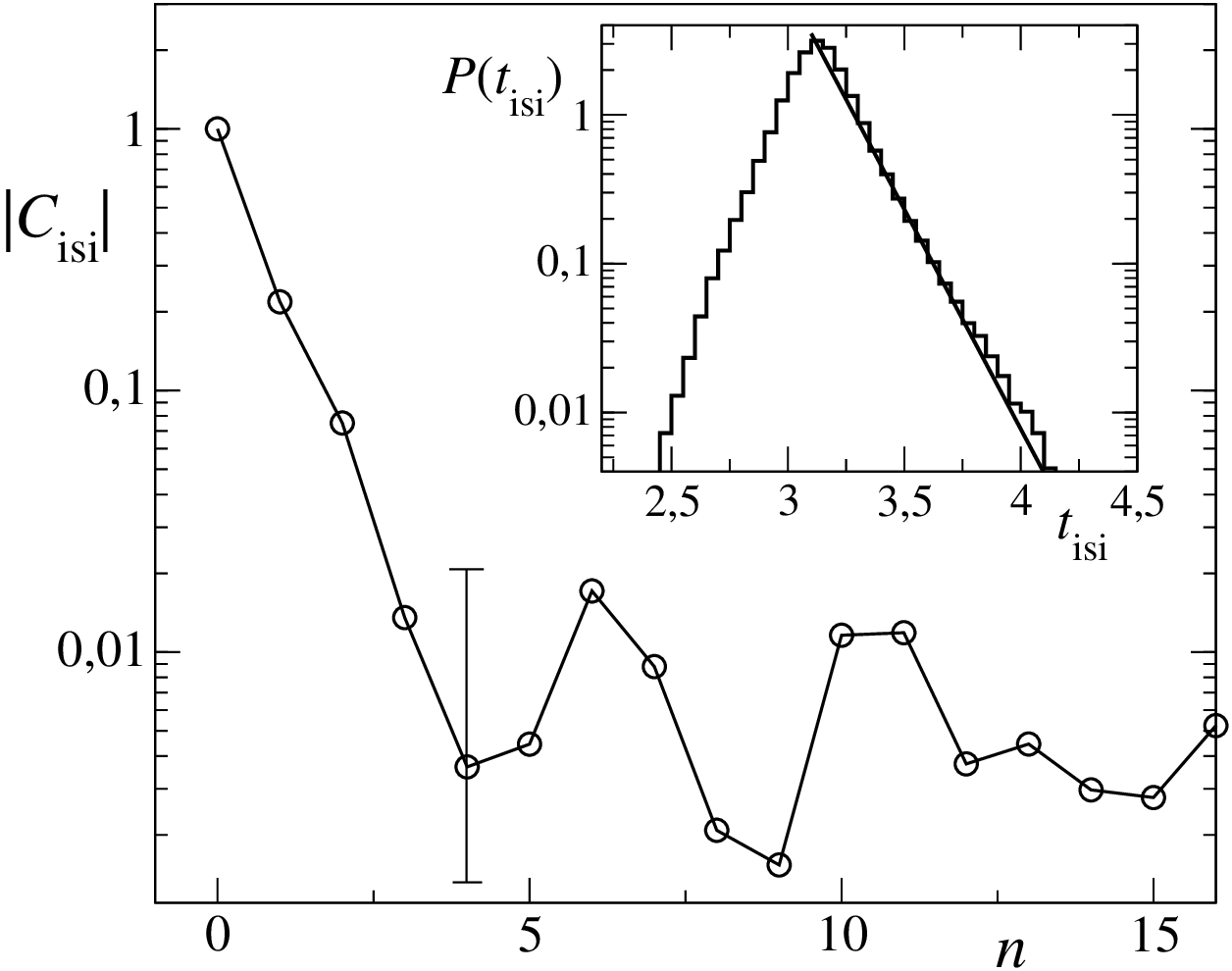}
\caption{The normalized autocorrelation $C_{\rm isi}$
  of the $t_{\rm isi}$ during transient
  for a single neuron {\em vs.} the integer delay for $N=1000 , G_0=2\,$.
  The inset shows the corresponding distribution $P(t_{\rm isi})$ that
  exhibits an exponential decay.}
\label{fig:corr}
\end{figure}

At higher coupling strengths ($G_0\agt 1\,$~) a different network dynamics
sets in. The ISI is no-longer constant and the resulting
pattern keeps slowly changing (e.g., see Fig.~\ref{fig:perpat1}).
We accordingly call this dynamical regime {\sl Unlocked Phase} (UP)~. The
phase variables corresponding to the different neurons evolve like independent
random walks, despite the dynamics is characterized by a negative maximum
Lyapunov exponent. One can understand the new scenario by observing that the
increased (inhibitory) coupling strength destabilizes (or just makes 
disappear) an increasing number of periodic spiking sequences. Accordingly, we
observe a continuous rearrangement of the ordering, until one of the
remaining stable sequences is reached. It is certainly desirable to understand
the rules of this pruning of periodic solutions, but we first prefer to
concentrate our efforts on the statistical characterization of the transient
regime, since the irregularity of the evolution can neither be traced back to
an exponential separation of orbits, nor to the action of an external noise
source.
We start by computing the normalized autocorrelation function of the ISIs,
$t_{\rm isi}(n)$, of the single neurons,
\[
  C_{\rm isi}(n)=\frac{\av{t_{\rm isi}(n')t_{\rm isi}(n'+n)}-
  \av{t_{\rm isi}}^2}{\av{t^{2}_{\rm isi}}-\av{t_{\rm isi}}^2}\ ;
\]
where $\av{\cdot}$ denotes the average over the time index $n'\,$.
$C_{\rm isi}$ (see Fig.~\ref{fig:corr}) clearly exhibits an exponential decay,
while the corresponding ISI distribution
(see the inset in the same figure) exhibits an exponential tail that is typical
of Poisson processes and testifies to the absence of correlations at long
times. 
\comment{
The existence of an exponential tail in the ISI distribution deserves a
specific discussion. In fact, an arbitrarily long ISI requires, in this context,
the possibility of a sufficiently strong inhibition to keep the membrane
potential away from the threshold. Indeed,
in trying to find a meaningful signature of the LP-UP transition, we
have discovered that it is only in the UP that the effective value of
$\dot{v}_{i}$ (after including the average effect of the received spikes)
can be negative. Therefore, the existence of exponentially long transients seems
to be connected to the possibility of having arbitrarily long ISIs.}

It is also important to test the stationarity of the UP. By computing the
coefficient of variation, $C_{\rm V}(n)\,$ (the standard deviation normalized to
the mean value) of the ISI in different windows, it can be
seen that while $C_{\rm V}(n)$ slowly decreases towards 0 in the LP (indicating
a convergence towards an ordered state), it remains constant in the UP
\cite{zillmer} although the asymptotic value is quite small.
For instance, we find that $C_{\rm V}\approx 0.03$ for $G_0=1.6$ in contrast to
values around 1, typically found for randomly spiking neurons in the
cortex \cite{Softky}. This can be due to the smallness of the disorder that is
considered in the network. However, it is also interesting to notice that
even in the present context for a different setup (i.e., under the action
of a periodic input) a much wider distribution can be found as discussed in
the next section.

\section{Response to a periodic signal}\label{sec:signal}

An interesting question is whether the emergence of long irregular transients 
in the presence of multiple periodic attractors allows performing
neurocomputational tasks. As a first test in this direction
we study the effect of a periodic signal on a subset $N'$ of the $N$
neurons. The signal is assumed to be an equispaced spike train with an 
excitatory action on the receiving neurons. The dynamics
of the unforced $N-N'$ neurons is still given by Eq.~\eqref{eq:model1}, 
while the remaining $N'$ neurons are subject to the extra excitatory
synaptic input,
\bey\label{eq:modelex}
  I_e= -(v_{i'}+w_{\rm e})G_{\rm e}\,\sum_{n=1}^{M_{\rm e}}\delta(t-t_{\rm e}^{(n)}) \ ;
\eey 
where $w_{\rm e}$ is the reversal potential, $G_{\rm e}\,$ represents
the coupling constant and $M_{\rm e}$ counts the external spikes 
received up to the time $t\,$. The ISI of the excitatory
post-synaptic potentials (PSPs) is assumed to be constant
(i.e., $t_{\rm e}^{(n+1)}-t_{\rm e}^{(n)}\equiv T_{\rm e}\,$). Thus the
effective signal strength is given by the ratio $G_{\rm e}/T_{\rm e}\,$.

\begin{figure}[h]
\begin{center}
(a)\includegraphics[width=0.37\textwidth,clip=true]
{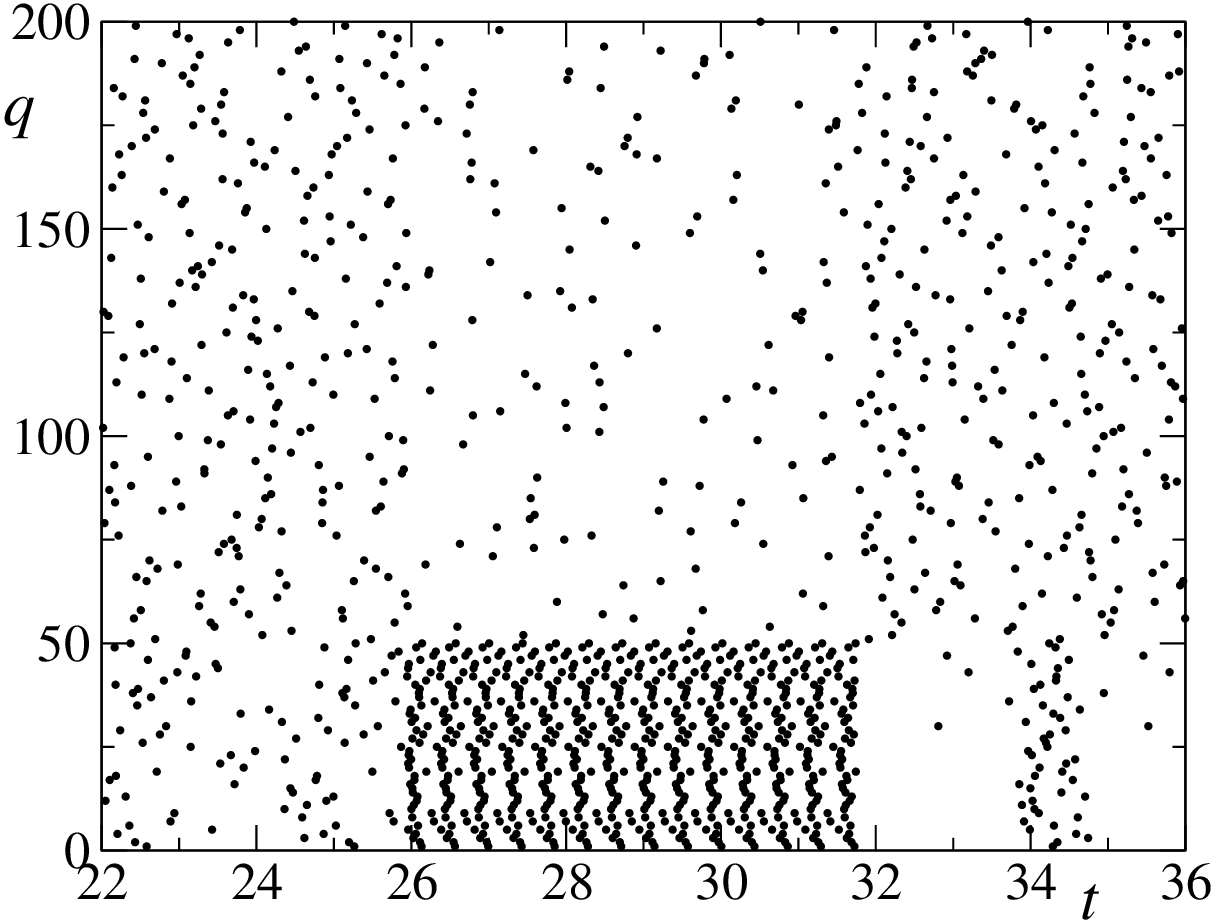}\hspace{7mm}
(b)\includegraphics[width=0.37\textwidth,clip=true]
{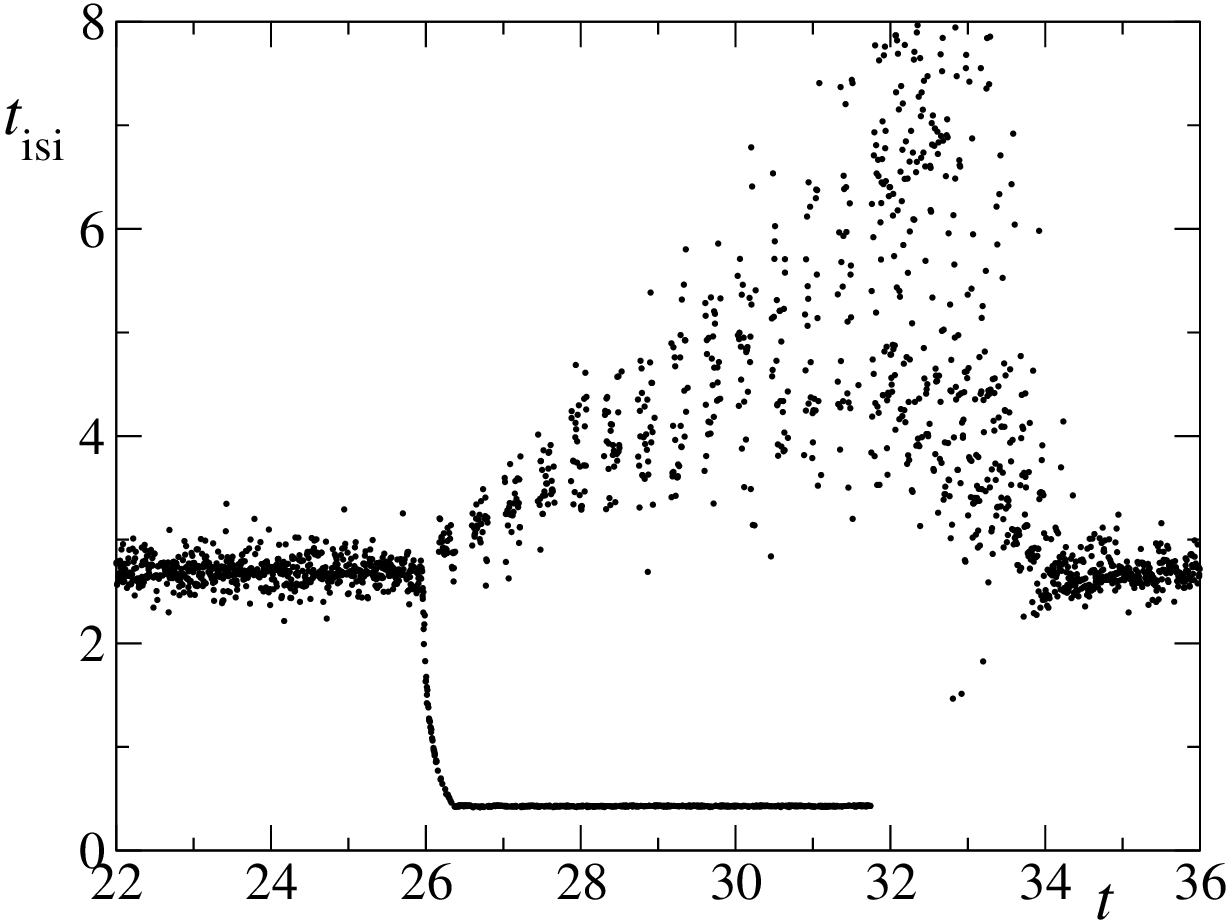}
\caption{(a) Firing pattern for $G_0=1.8,N=500\,$; $q$ denotes the index of the 
firing neuron. The excitatory input was appplied to neurons $i=1,\cdots 50$
for $t\in [25.9,31.8]\,$. To enhance visibility only the first 200 neurons
are shown.
(b) The $t_{\rm isi}$ of the respective firing neuron. With the applied signal
the $t_{\rm isi}$ of the receiving neurons lock to $t'_{\rm isi}=0.43\,$.}
\label{fig:signal}
\end{center}
\end{figure}

Due to the excitatory nature of the PSP, the firing rate of the $N'$ neurons
increases, and this, in turn, gives rise to an increased inhibitory activity in
the network. Thus the activity of the $N-N'$ ``background'' neurons that do not
partake in the excitatory input should diminish compared to the subset $N'\,$.
In Fig.~\ref{fig:signal} we present results for a system of $N=500$ neurons
in the UP ($G_0=1.8$). The input signal is switched on during the time interval
$t\in [25.9,31.8]\,$; otherwise the network evolves freely according to
Eq.~\eqref{eq:model1}. For the PSP input trains, we have taken the values
$w_{\rm e}=-8$, $T_{\rm e}=10^{-3}$ and $G_{\rm e}=2.2\times 10^{-4}\,$, i.e., 
$G_{\rm e}/T_{\rm e}=0.22\,$. The input is applied to the first 50 neurons
(given the equivalence on average among all neurons, the choice of the label is
totally irrelevant);  accordingly $N'/N=0.1\,$. By looking at the firing pattern
shown in Fig.~\ref{fig:signal}(a), one can clearly notice the effect of the
excitatory input from the significantly increased firing frequency of the 
driven neurons, while, simultaneously, the activity of the undriven
background neurons diminishes. Not surprisingly, the pattern of the $N'$
neurons is much more regular than the spike pattern of the background. This is
better seen by looking at the ISI's, $t'_{\rm isi}\,$, of the
driven neurons in Fig.~\ref{fig:signal}(b). Indeed, soon after the input has been
switched on the $t'_{\rm isi}$ rapidly decrease towards $0.43$ with small
fluctuations $< 2\,\%\,$. Once the signal has been switched off, the system
rapidly returns to the initial irregular pattern after the neurons have been
reset. 

When the input signal is on, the forced neurons behave almost periodically
with a period $t'_{\rm isi}\gg T_{\rm e}\,$. The small fluctuations are
due to the inhibitory action of the background neurons. However,
the negative Lyapunov exponent ensures stability with respect to small 
perturbations and thereby guarantees a robust synchronization of the subset.
The $N'$ neurons somehow resemble the ``active'' population of cells
that respond to a certain stimulus by firing with an increased rate.

On the other hand, the ISIs of the background are more
irregular than in the freely evolving network as revealed by the strongly enhanced
fluctuations. In order to perform a more accurate description of
their behavior, we have switched on the input for longer times, so as to be able
to determine the probability distribution of the intervals $t_{\rm isi}$ in the
stationary regime. 
\begin{figure}[h]
\begin{center}
(a)\includegraphics[width=0.31\textwidth,clip=true]
{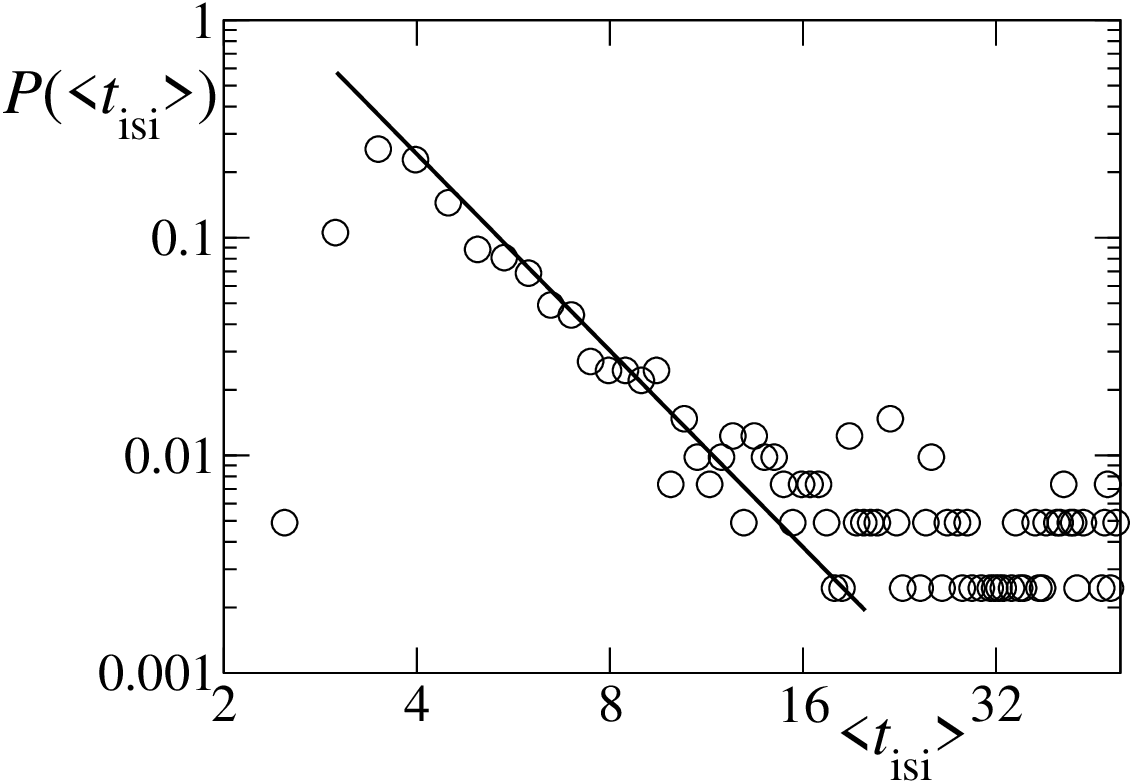}
(b)\includegraphics[width=0.35\textwidth,clip=true]{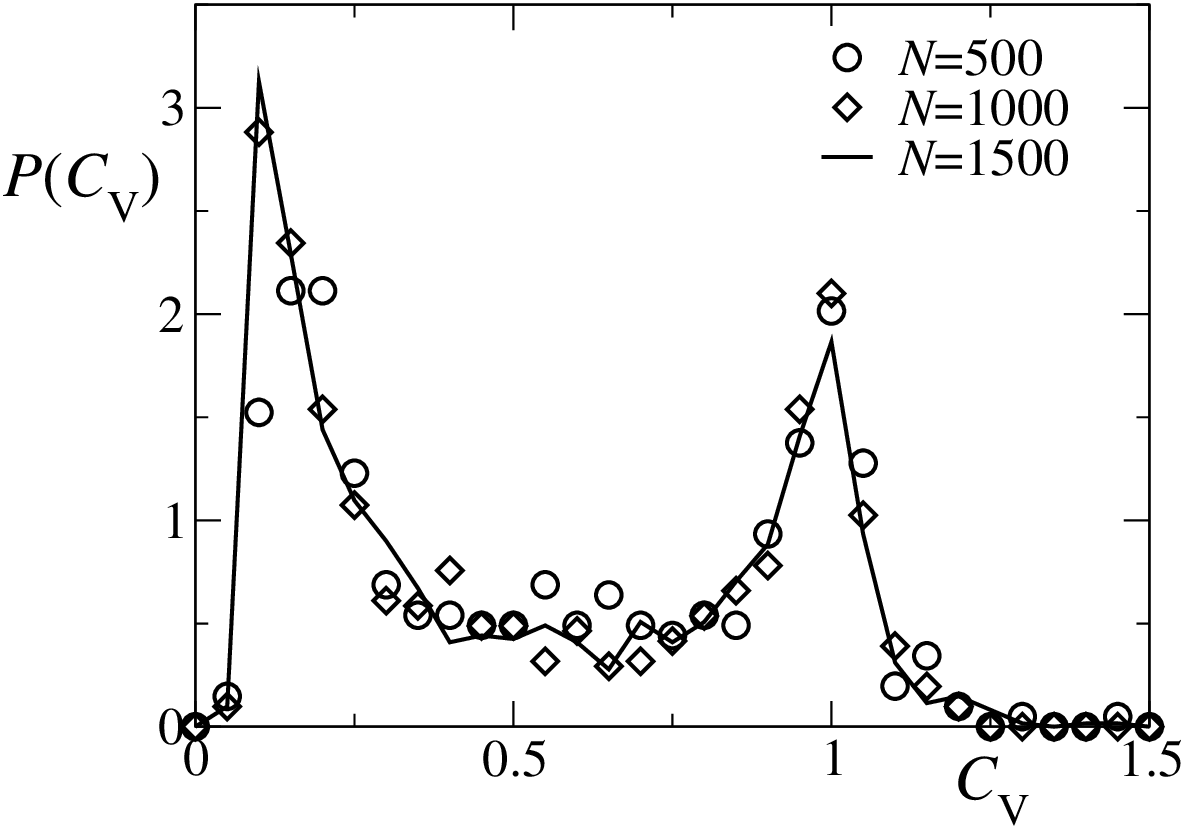}
(c)\includegraphics[width=0.3\textwidth,clip=true]{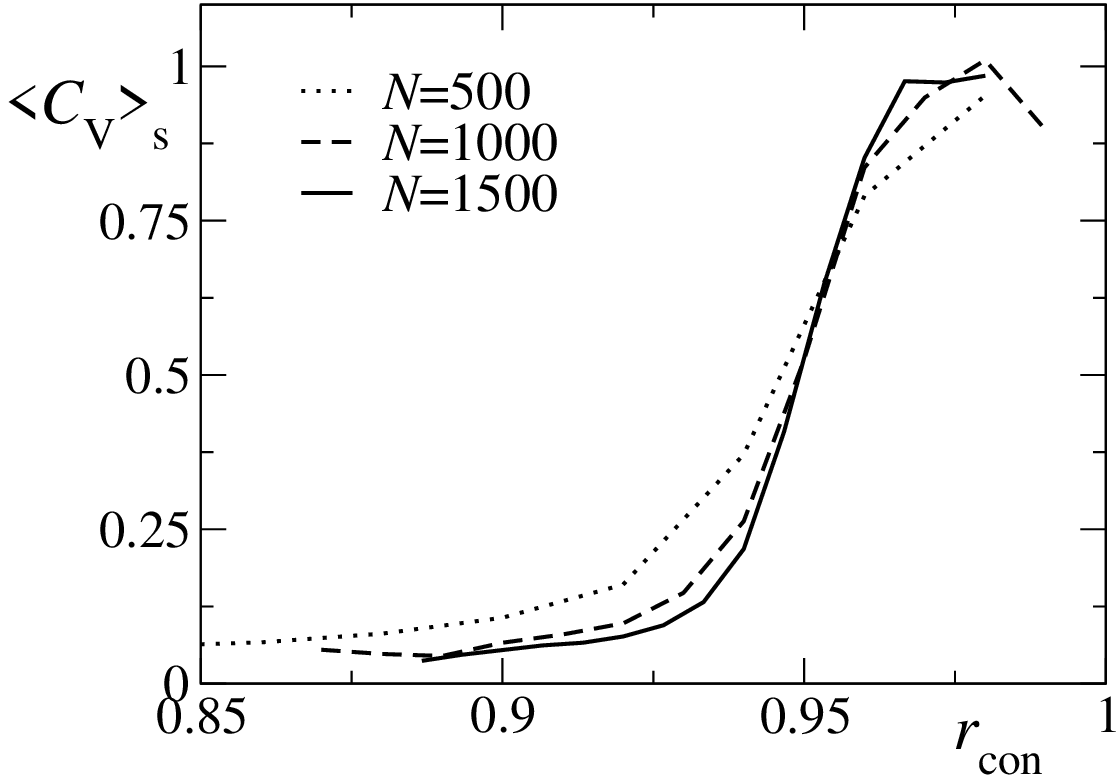}
\caption{(a) The stationary distribution of the ``background'' $\av{t_{\rm isi}}$ 
for the same parameter values as in Fig.~\ref{fig:signal} (circles). The solid line 
is a fit with a power-law with exponent $3$.
(b) The stationary distribution of the $C_{\rm V}$ of single neurons
belonging to the ``background'' for different system sizes ($N'/N=0.1$).
(c) The average $\av{C_{\rm V}}_{\rm s}$ as a function of the
connectivity $r_{\rm con}$ for different system sizes ($N'/N=0.1$).}
\label{fig:poissdis}
\end{center}
\end{figure}
It turns out that different background neurons show substantial differences
in their firing rate, $\av{t_{\rm isi}}^{-1}\,$. In Fig.~\ref{fig:poissdis}(a)
the distribution of $\av{t_{\rm isi}}$ is shown, that can be fitted for intermediate
values by a power-law decay with an exponent $\approx 3$. However, the tail
shows significant deviations from the power-law, which suggests a decomposition
of the background neurons into different classes. 
In order to get a deeper insigth into this phenomenon, we
have examined the distribution of single neuron $C_{\rm V}$'s for different
system sizes keeping $N'/N=0.1$ (see Fig.~\ref{fig:poissdis}(b)). As a matter of fact, 
this characterization of the firing processes reveals a bimodal
distribution. The peak seen at smaller values is
reminiscent of the behaviour of the freely evolving network: we can claim
that such neurons are almost unaffected by the input signal. On the other hand,
the neurons with $C_ {\rm V}\simeq 1$ exhibit a Poissonian ISI distribution with a
much more pronounced exponential tail. The different behavior of background neurons
can be understood by examining for each neuron the number of intact incoming links 
from forced neurons, that after a normalization with $N'$ yields the
local connectivity $r_{\rm con}\in [0,1]\,$. A large value of $r_{\rm con}$ implies 
an enhanced inhibition of the respective neuron.
We collected neurons that share the same
value of $r_{\rm con}$ and computed the corresponding average $C_{\rm V}$
that we denote as $\av{C_{\rm V}}_{\rm s}\,$. Indeed, the results in
Fig.~\ref{fig:poissdis}(c) show that with increasing
$r_{\rm con}$ there is a transition-like behavior of $\av{C_{\rm V}}_{\rm s}\,$,
as it rapidly increases from values around 0.05 to values around 1\,.

%%%%%%%%%%%%%%%%%%%%%%%%%%%%%%%%%%%%%%%%%%%%%%%%%%%%%%%%%%%%%%%%%%%%%%%%%%%%%
\section{Conclusions and perspectives}
We have studied a diluted inhibitory network of pulse-coupled neurons that is
{\it dynamically stable} in the whole parameter range considered. 
The dynamical stability, determined by a negative Lyapunov exponent,
ensures the existence of asymptotic periodic attractors.
By varying the
coupling strength $G_0$, a desynchronization transition is found:
for $G_0\agt 1.2$ the dynamics is governed by {\it chaotic-like transients}
whose duration grows exponentially with the system size.
The transients have good stationarity properties and they are characterized by
irregular firing sequences with an exponentially tailed ISI distribution. Such a
behavior is usually attributed to dynamically unstable
chaotic networks, e.g., with balanced excitatory and inhibitory activity
\cite{Vreeswijk-Sompolinsky:1996}. 
With the help of mean field arguments, the basic mechanism of the
desynchronization can be identified: the dropout of input due to a pruned link
which leads to irregular changes of the neurons position in the firing sequence.
Thus we expect the effect to emerge in a wider class of models (e.g.~inhibitory
networks with small fractions of excitatory links). Preliminary simulations
have further shown the effect to persist in the presence of delayed interaction.

We want to point out that only under the peculiar conditions of
{\sl stable chaos} (SC) it is possible to reconcile linear stability with an
unpredictable, chaotic evolution: as soon as the system size is sufficiently
large, the periodic attractor is practically unreachable and the erratic
transient dynamics is the only accessible information from the system. 
These findings could be of some relevance in connection with the recent experimental
results of Mazor and Laurent \cite{mazor}, who pointed out that odor discrimination
is better performed by projection neurons in the locust antennal lobe
during transient dynamical phases preceeding stable states.

In biological systems, the effect of noise has to be taken into account. In 
the present case this could manifest itself as external background noise or as
internal fluctuations such as synaptic failures. For small enough noise
amplitudes, the linear stability of the considered system ensures robustness
of the presented behavior. We have shown this by synchronizing a subset of
the neurons, where the irregular activity of the other neurons can be seen as 
weak background noise. In the case of larger noise, one has to consider
two cases. In the desynchronized phase ($G_0\agt 1.2$) the noise amplifies
the already present irregularities which renders the convergence towards a
periodic attractor impossible. For what concerns the locked phase
($G_0\alt 1.2$), we found that finite perturbations switch the dynamics between
the different periodic attractors. However, a detailed investigation of noise
effects is still in progress.

The observed mechanisms provided by SC should be present in
a dynamical model of a system which is expected to reliably perform 
universal computations. Indeed, thanks to its asympotic stability, the SC
regime is found to be naturally stable with respect to external
perturbations. On the other hand, the long and stationary transient ensures that
the system can explore an extremely rich variety of configurations.

This gives rise to the challenging question, whether it is possible to address
the various periodic states in a reproducible way by properly chosen inputs. In this
spirit a study of the properties of the considered dynamics including external
drives (information input) is of high interest. As a first test, we have
demonstrated that it is possible to lock by external periodic signals a
relatively small subset of 
neurons onto a periodic attractor while the rest (background) of the network 
performs an irregular activity. Interestingly, the background decomposes into
two classes with respect to the $C_{\rm V}$ of the single neuron ISI distribution,
where one class is characterized by a $C_{\rm V}$ close to unity.
All of these remarks indicate that the locked as well as the
desynchronized phase of our model may contain ingredients 
of direct interest for applications to neurosciences.

Finally, we wish to emphasize that from a conceptual point of view, it is
remarkable that SC has been found in a continuous-time autonomous dynamical
system such as model (\ref{eq:model1}).  After its
discovery in space- and time-discrete models (coupled map lattices)
\cite{stable}, the only known case of a continous-time dynamical model
exhibiting SC is a chain of periodically forced Duffing oscillators
\cite{bonac}. In fact, an autonomous dynamical system depending on
a continuous time variable is expected to be marginally stable, i.e.,
$\Lambda = 0$. Such a statement can be made rigorous as soon as the
interactions which determine the dynamics of such a system are assumed
to be smooth, differentiable functions of its variables. This is not the
case of our model, which is certainly continuous in time but incorporates
strong nonlinear, i.e., non-smooth, properties due to the threshold 
mechanism ruling firing events. It is this
constituent property, typical of neural dynamics, which essentially 
turns the continous time dynamics to a map-like behavior, thus rendering
possible the emergence of SC. 
%%%%%%%%%%%%%%%%%%%%%%%%%%%%%%%%%%%%%%%%%%%%%%%%%%%%%%%%%%%%%%%%%%%%%%%%%%%%%%
%       References
%%%%%%%%%%%%%%%%%%%%%%%%%%%%%%%%%%%%%%%%%%%%%%%%%%%%%%%%%%%%%%%%%%%%%%%%%%%%%%

\end{document}